# Comparative Analysis of GPGPU based ACO and PSO Algorithm for Employee Scheduling Problems

Harsha R. Gaikwad[1], Pradnyarani K. Mahind[2], Sandeep U. Mane[3]

[1]M. Tech. Student Department of CSE, Rajarambapu Institute of Technology
Islampur (Sangli), MS, India

[2]M. Tech. Student Department of CSE, Rajarambapu Institute of Technology
Islampur (Sangli), MS, India

[3]Assistant Professor Department of CSE, Rajarambapu Institute of Technology
Islampur (Sangli), MS, India

[1]harshagaikwad4@gmail.com
[2]mahindpradnya@gmail.com
[3]sandeep.mane@ritindia.edu

*Abstract:* Particle Swarm Optimization (PSO) and Ant Colony Optimization (ACO) are simple, easy to implement, its robustness to control parameters, and its computational efficiency when compared with mathematical algorithm and other heuristic optimization techniques. The calculation in PSO and ACO is very simple. Compared with the other developing calculations, it occupies the bigger optimization ability and it can be completed easily. It is used to solve many NP-Hard problems. Employee Scheduling is a real-life NP-Hard problem faced by many organizations. Self-scheduling in all situations is not always practical and possible. Nurse Rostering is related to highly constrained resource allocation problem into slots in a legal shift Earlier the problem was solved using different heuristic algorithms. In this dissertation, we have proposed, GPGPU based parallelization of PSO and ACO to solve Employee scheduling problems. To parallelize both algorithms, a master-slave approach is used. The BCV 8.13.1 data set is used for experimentation purposes. Analysis of results is done based on mean, standard deviation, standard mean error.

*Keywords: Employee Scheduling, Parallelization, PSO, GPGPU.*

## 1. Introduction

Scheduling problem is the resource allocation problem. Given a set of resources with some capacities, a set of activities with processing time and resource requirements, and a set of constraints between activities, a scheduling problem consists of deciding when to execute each activity, so that both temporal and resource constraints are satisfied (Le Pape,2005). Most scheduling problems can easily be represented as instances of the constraint satisfaction problem; given a set of variables, a set of possible values (domain) for each variable, and a set of constraints between the variables, assign a value to each variable, so that all the constraints are satisfied. Most scheduling problems can easily be represented as instances of the constraint satisfaction problem; given a set of variables, a set of possible values (domain) for each variable, and a set of constraints between the variables, assign a value to each variable, so that all the constraints are satisfied(Van den Bergh, 2013).

There are different types of scheduling problems like job shop scheduling problems in which N jobs (N1, …., Nn) of different sizes are given, which need to be scheduled on M identical machines the objective is to minimize the overall makespan. There are many variations of job shop scheduling problems such that machines can be related, of equal capacity or unequal capacity, machines require some time between executions of jobs, machines can have sequence-dependent setup, etc. Flow shop scheduling problem, flow shop scheduling problem is just like job shop scheduling problem the only difference is that the jobs need to be executed in specific processing order. In the Employee scheduling problem numbers of employees are assigned to the available shifts; to satisfy the constraints and requirements of the organization.

Employee Scheduling Problem Employee scheduling is a real-life frequently occurring optimization problem. Employee scheduling problem arises in many organizations including scheduling of nurses in the hospital; check encoders in the bank, airline, and hotel reservation personnel, telephone operators scheduling, etc. Employee scheduling is a resource allocation problem. The general employee scheduling problem is defined as the assigning number of employees to the available shifts or days to fulfill the requirements of the organization and the employees. Employee scheduling is a discrete constrained optimization problem (Blazewicz, 1994). The objective is to build a working timetable for each Employee by defining the start time of work, the end time of the work, the free time required to the Employee before the work starts and after the work ends, the total number of working days for each Employee per week, etc.

Following covers details of instance we have implemented

A. Specifications of instance 8.13.1

1) *Number of Nurses -13*

2) *Number of Days - 28*

3) *Number of shifts - 5*

There are 5 different shifts are included in the given benchmark dataset, the shifts are described as follows:

**Table 1. Dataset shifts description**

| Shifts | Start Time | End Time | Description |
|---|---|---|---|
| V | 06:00 | 14:00 | Vroege |
| D | 08:00 | 17:00 | Dagdienst |
| DH € | 08:00 | 17:00 | Dagdienst (Head Nurse) |
| L | 14:00 | 22:00 | Late |
| N | 22:00 | 06:00 | Night shift |

4) *Hard Constraints*

Hard constraints are must be satisfied, Hard constraints is the condition which must be true.

**Table 2. Hard Constraints**

| HC1 | Nurse should not have 3 consecutive night shifts. |
|---|---|
| HC2 | At most one shift is assigned per day per nurse |
| HC3 | There should not be more than 5 blank shifts per day |
| HC4 | Nurse must match the skills required for that shift. |

5) *Soft Constraints*

Soft constraints are the condition that may or may not be true, each soft constraint is having a specific weight and some penalty cost for its violation. Several times the constraint violates will increase the penalty cost

**Table 3. Soft Constraints (http:// www.cs.nott.ac.uk/tec/NRP/, 2014)**

| Soft Constraints | Description | Weight |
|---|---|---|
| SC1 | Max consecutive free days | 0 |
| SC2 | Max hours Worked | 1 |
| SC3 | Complete weekends | 40 |
| SC4 | Max shift types | 5 |
| SC5 | Requested shifts on | 0 |
| SC6 | Number of consecutive shift types | 5 |
| SC7 | Requested shifts Off | 0 |
| SC8 | Maximum consecutive working days | 5 |
| SC9 | Max Shift types per week | 1 |
| SC10 | Requested days off | 0 |
| SC11 | Shift type successions | 10 |
| SC12 | Max shifts for days of the week | 1 |
| SC13 | Skilled shifts | 10000 |
| SC14 | Alternative skill | 1 |
| SC15 | Min time between shifts | 20 |
| SC16 | Night shifts Before free weekend | 0 |
| SC17 | Max working bank holidays | 10 |
| SC18 | Max number of shifts | 5 |
| SC19 | Min consecutive free days | 1 |
| SC20 | Max working weekends in 4 weeks | 7 |
| SC21 | Min Consecutive working Days | 1 |

**2. Implementation Details**

We have solved employee scheduling problem using both ant colony optimization and particle swarm optimization on GPGPU. In literature, many scheduling problems are solved using both algorithms.

In (Deng, 2012) an airline crew scheduling is an optimization problem, an effective airline crew schedule reduces the cost of the airline industry. Airline Crew scheduling is assigning the members of the crew to the flight for a specific period. Ant Colony Optimization is inspired by the natural behavior of the real ants; as ACO is successfully applied to the traveling problem so, in this paper (Deng, 2012), the airline crew scheduling problem is modeled as TSP problem and then the ACO algorithm is applied.ACO is proposed to solve airline crew scheduling problems by applying the flight-based scheduling representation and trying to find the shortest path from flight graph like the Travelling salesman problem. ACO is a powerful and robust algorithm when the results getting from ACO are compared with GA it seems that ACO gives better performance for airline crew scheduling problems.

In (Gutjahr, 2008), describes the first ACO applied to the dynamic Nurse Scheduling problem. Everyday group of nurses is assigned to the public hospitals, taking into the account number of hard and soft constraints like date and time of work, working pattern, qualification of nurses, number of nurses and number of hospitals and cost. Solving this problem manually is time and cost-consuming. To solve the problem author proposed the ant Colony Optimization algorithm, motivated by the fact that ACO has given better results for many scheduling problems like vehicle routing, job shop scheduling. The results of the simulation show that ACO gives better performance than the other greedy algorithms.

A. Working Of ACO

Ant Colony optimization is the probabilistic algorithm invented in 1992 by Marco Dorigo.Marco Dorigo first applied the Ant colony optimization algorithm to solve the traveling salesman problem. An algorithm is inspired by the natural behavior of ants; when ants explore the paths when they search the food. ACO is an evolutionary algorithm with many advantages. In the general working of Ant Colony optimization while finding the food source each ant randomly chooses the path for food source during the traveling from an initial position to the food source ant deposit a chemical substance i.e. pheromone on the path, based on the amount of pheromone deposited on the path other ants choose the path. In the process of creating a path each ant update the pheromone of the edges that have gone through according to the local updating rule (pei, 2012).

Pseudo-code for Ant Colony Optimization
1. Create construction graph
2. Initialize pheromone values
3. While not stop-condition do
4. Create all ants' solutions
5. Perform local search
6. Update pheromone values

B. Variations of Ant Colony Optimization

1) *Basic Ant System*

There are many variations of Ant colony optimization depending on the pheromone update method of the ants. In the basic ant system, pheromone values are updated by all m ants that have built their solution in iteration itself.

2) *Ant Colony System*

In the Ant Colony system pheromone update is taking place at the end of each construction process in addition to this local pheromone update is taking place. Local pheromone update is taking place by each ant after is construction step of each construction process.

3) *MAX MIN Ant System*

In the MAX MIN ant system only the best ant i.e. the ant who is getting the best solution at the end of each iteration update the pheromone and that value is the maximum bound on the pheromone the main focus of this method is on the iteration best solution (Marco Dorigo,2006).

4) *Rank Based ACO*

In the ant system with ranking, suppose n ants have built the solutions, so the ants are sorted based on good solutions they have found the rank is given to each ant after sorting and the contribution of ant in the pheromone update based on their rank (Marco Dorigo,2006). The stochastic component of ACO allows the ants to build different solutions and allows exploring the search than greedy heuristic (Marco Dorigo, 2006).

5) *Elitist Strategy of ACO*

The emphasis is given on the best solution found so far in the elitist strategy of ACO (Marco Dorigo, 2005).At every iteration worst solution is replaced by the best solution. This is the most popular and widely used approach of Ant Colony Optimization.

C. Working Of PSO

Particle Swarm Optimization (PSO) is a population-based optimization technique, inspired by the social behavior of bird flocking or fish schooling. In PSO each particle is a candidate solution that flows through the search space, adjusting their position by their own best position or concerning its neighbor's best position towards an optimal solution. The performance of every particle is checked by the fitness function. At each iteration, particle updates their position and velocity while moving towards an optimal solution. PSO is sensitive with its parameters like acceleration coefficients and inertia weight (Blondin, 2009).

Working of PSO is as follows:-
1. Initialize a population of particles with random positions and velocities on d dimensions in the problem space.
2. For each particle, evaluate the desired optimization Fitness function.
3. Compare particle's fitness evaluation with particle's pbest.
4. If the current value is better than pbest, then set pbest value equal to the current value and the pbest location equal to the current location in d-dimensional space.
5. Compare fitness evaluation with the population's overall previous best.
6. If the current value is better than gbest, then reset gbest
   to the current particles array index and value.
7. Change the velocity and position of the particle.
   According to equations (1) and (2), respectively:

$$v_{id} = v_{id} + c_1 * rand1() * (p_{id} - x_{id}) + c_2 * rand2() * (p_{gd} - x_{id})$$
(1)

$$x_{id} = x_{id} + v_{id}$$
(2)

D. Variations of PSO

**Table 4. Variations of PSO (Rini, 2011)**

| Sr. No. | Basic Variant | Function | Advantages | Disadvantages |
|---|---|---|---|---|
| 1. | Velocity Clamping | Global exploration of the particle is controlled. A particle remains in | Movement of the particle is controlled due to reduced step velocity. | If all the velocity becomes equal to the particle will continue to conduct |

| | | the search space because it reduces the step velocity. It cannot change the search direction of the particle | | searches within a hypercube and will probably remain in the optima but will not converge in the local area. |
|---|---|---|---|---|
| 2. | Inertia Weight | It takes into account the previous velocity and controls the momentum of the particle. | A larger inertia weight at the end of the search will foster the convergence ability. | Achieve optimality convergence strongly influenced by the inertia weight |
| 3. | Constriction Coefficient | To ensure the stable convergence of the PSO algorithm | Similar with inertia weight | When the algorithm converges, the fixed values of the parameters might cause the unnecessary swinging of particles |
| 4. | Synchronous and Asynchronous Update | Optimization in parallel processing | Improved convergence rate | Higher throughput, Higher accuracy |

E.  General Purpose Graphics Processing Unit (GPGPU)

GPU(Graphics Processing Unit) accelerated computing is the use of a graphics processing unit (GPU) together with a Central Processing Unit to speed-up scientific, analytics, engineering, consumer, and enterprise applications. Excellent and never achieved performance can be acquired by shifting compute-intensive portions of the application to the Graphics Processing Unit for execution and executing the rest portions of the application on the Central Processing Unit. From the user's point of view, the difference is just that the application runs notably faster. A GPU consists of thousands of small cores while a CPU has few cores for sequential processing (http://www.nvidia.com/object/, 2014).

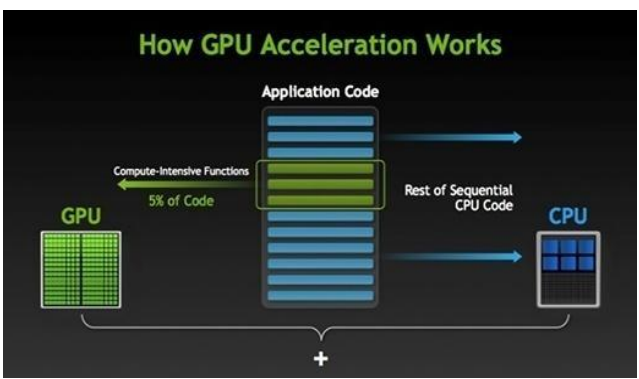

**Fig. 1 Working of GPU**

### 3.  Result and Discussion

We have implemented CUDA based parallel Ant Colony Optimization algorithm and Particle Swarm Optimization Algorithm to solve problem instance 8.13.1. Ant Colony optimization is inherently parallel; we have implemented GPGPU based parallel Ant Colony optimization to solve the Employee Scheduling problem. The algorithm-specific parameters of Ant Colony optimization, α, β, η, ς kept constant; where η is problem-specific heuristic information, while ς is Pheromone evaporation rate. Experimental results show that GPGPU based parallel Ant colony optimization gives a better solution. In the perspective of GPGPU, as we have increased the number of ants, utilization of several blocks and the number of threads also increased. The parameter settings areas,

$$\alpha=1$$
$$\beta=2$$
$$\eta=0.01$$
$$\varsigma=0.5$$

In the case of Particle swarm Optimization we have varied population as16, 32 and c1=c2 (construction factor) varied as 1.2 and 1.5. It is observed that for population 32 with c1=c2=1.5 better results are obtained as compared to population 16 and c1=c2=1.5. It is observed that for population 32 with w=0.9 better results are obtained as compared to population 16 and w=0.9. We have concluded that at c1=c2=1.5 w=0.9 better results are obtained. The population is set 16. When we set algorithm-specific parameters c1=c2=1.2 and w=0.4 and experimental result shows that as the number of generations increases fitness value decreases and the fitness value is near to the optimal solution. When we set algorithm-specific parameters c1=c2=1.2 and w=0.9 and experimental result shows that as the number of generations increases fitness value is also increasing. It shows as we are changing w (inertia weight) from 0.4 to 0.9 it is not giving optimal results for 16 particles. Algorithm specific parameters c1=c2=1.5 and w=0.9 and experimental results show that several generations increase fitness value decreases and fitness value obtained is near to optimal solution.

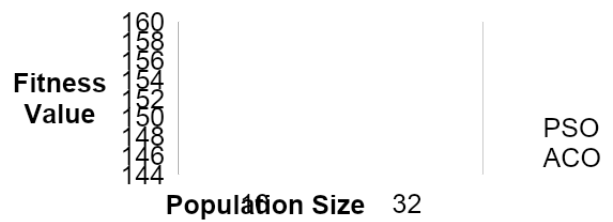

**Fig. 2 Population Size vs. Fitness value for the number of iterations =1000**

Fig. 2 and Fig. 3 give the comparative study of Parallel Ant Colony Optimization and particle swarm optimization.

Graph 1 is for population size versus the number of iterations. As we increase the population size fitness value increases. We are proceeding to global optimal as population size is increased in both ant colony optimization and particle swarm optimization.

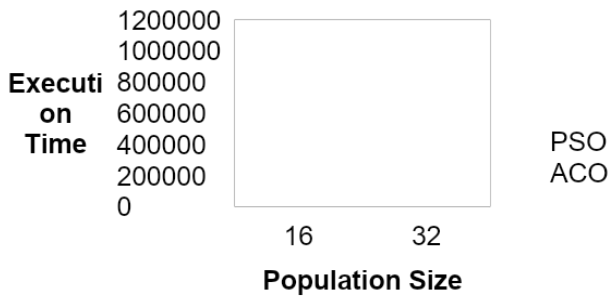

Fig 3. Execution Time Vs. population size for several iterations =1000

Fig. 3 is execution time versus population size as we increase population size we get better results and execution time increases. For Ant Colony Optimization drastic change in time is observed as population size is increased whereas a slight change in time is observed in Particle swarm optimization when the population is varied.

4. **Conclusion**

Ant colony optimization and Particle Swarm Optimization are heuristic techniques that are inspired by the natural behavior of ants and Birds. In literature, ACO and PSO have been successfully applied to many NP-hard problems. It gives better results for routing problems, TSP, and scheduling problems. As a population-based algorithm, it is intrinsically parallel, and thus well suited to implement on parallel architectures. We have implemented a master-slave approach of parallelization of Ant Colony Optimization and Particle swarm optimization to solve the Employee Scheduling problem on GPGPU. For experimentation purposes, the BCV 8.13.1 data set is used. Experimental results show that parallelization of ACO gives better results than that of PSO.